\documentclass[a4paper]{jpconf}
\usepackage[pdftex]{graphicx}
\usepackage{amsmath}
\usepackage{amssymb}
\usepackage{color}
\newcommand{\flabel}[1]{\label{f:#1}}

\newcommand{\fig}[1]{Fig.~\ref{f:#1}}

\newcommand{\sect}[1]{Section~\ref{s:#1}}

\renewcommand\vec[1]{\boldsymbol{#1}}
\newcommand{\mean}[1]{\left\langle #1 \right\rangle}
\newcommand{\elabel}[1]{\label{e:#1}}
\newcommand{\slabel}[1]{\label{s:#1}}
\newcommand{\eq}[1]{Eq.~(\ref{e:#1})}
\newcommand{\quot}[1]{``#1''}
\newcommand{\eqntext}[1]{ \left\{\!\!\!\mbox{\begin{tabular}{c}  #1
\end{tabular}}\!\!\! \right\}}
\newcommand{\xvec}{\vec{r}}
\newcommand{\alltheX}{X}
\newcommand{\deltavec}{\vec{\delta}}
\newcommand{\ellvec}{\vec{\ell}}
\newcommand{\elldir}{\ensuremath{\vec{e}_{\ell}}}

\newcommand{\cfigure}[3][\columnwidth]{
\begin{figure}[!htb]
\begin{center}
  \includegraphics[width=#1]{Figures/#2}\hspace{2pc}%
  \caption{#3}
  \label{f:#2}
\end{center}
\end{figure}}
\begin{document}

\newcommand{\sfigure}[3][14pc]{
\begin{figure}[h]
\begin{minipage}[b]{14pc}
   \includegraphics[width=#1]{Figures/#2}\hspace{2pc}%
\end{minipage}
\begin{minipage}[b]{14pc}\caption{#3}ar 
\label{f:#2}
\end{minipage}
\end{figure}}

\title{Sampling from a polytope and hard-disk Monte Carlo}

\author{Sebastian C. Kapfer and Werner Krauth}

\ead{sebastian.kapfer@ens.fr, werner.krauth@ens.fr}

\address{
   Laboratoire de Physique Statistique, Ecole Normale Sup\'erieure,
   UPMC, CNRS\\ 24 rue Lhomond, 75231 Paris Cedex 05, France}

\begin{abstract}
The hard-disk problem, the statics and the dynamics of equal two-dimensional
hard spheres in a periodic box, has had a profound influence on statistical and
computational physics. Markov-chain Monte Carlo
and molecular dynamics were first discussed for this model. Here we reformulate
hard-disk Monte Carlo algorithms in terms of another classic problem, namely the
sampling from a polytope. Local Markov-chain Monte Carlo, as
proposed by Metropolis et al.~in 1953, appears as a sequence of random walks in
high-dimensional polytopes, while the moves of the more powerful event-chain
algorithm correspond to molecular dynamics evolution.
We determine the convergence properties of Monte Carlo methods in a
special invariant polytope associated with hard-disk configurations, and the
implications for convergence of hard-disk sampling.  Finally, we discuss
parallelization strategies for event-chain Monte Carlo and present
results for a multicore implementation.
\end{abstract}

\section{Introduction}
The hard-disk system is a fundamental model of statistical and computational
physics. During more than a century, the model and its generalization to
$d$-dimensional spheres have been central to many advances in
physics. The virial
expansion is an example: Boltzmann's early calculations of the fourth virial
coefficient~\cite{Boltzmann_1896} ultimately led to Lebowitz and Onsager's
proof of the convergence of the virial expansion up to finite 
densities~\cite{Lebowitz_1964} for all $d$ and to the general and systematic
study of
virial coefficients. The
theory of phase transitions provides another example for the lasting
influence of the hard-disk model and its generalizations. Kirkwood and
Monroe~\cite{Kirkwood_1940} first
hinted at the possibility of a liquid--solid transition in three-dimensional
hard spheres. This prediction was surprising because of the absence of 
attractive interactions in this system.
The depletion mechanism responsible for the effective-medium attraction was also
first studied in hard spheres, by Asakura and Oosawa~\cite{Asakura_1954}. In
two dimensions, the liquid--solid phase transition
was first evidenced by Alder and Wainwright~\cite{Alder_1962}. It lead
to far-reaching theoretical~\cite{KTHNY}, computational~\cite{Mak_2006,Bernard_2011} and
experimental~\cite{Zahn_1999}
work towards the understanding of 2D melting. In
mathematics, hard disks and hard spheres have also been at the center of
attention~\cite{Diaconis_2010}.  A rigorous existence proof of the 
melting transition in hard spheres is still lacking, but
the ergodicity of the molecular dynamics evolution of this system
has now been established rigorously~\cite{Sinai_1970,Simanyi_2003}.

Arguably the most important role for the hard-disk model has
been in the development of numerical simulation methods. Molecular
dynamics~\cite{Alder_1957,Rapaport_1980} and 
Markov-chain Monte Carlo~\cite{Metropolis_1953} were first formulated for hard
disks.
The early algorithms have continued to be refined: Within the 
molecular dynamics framework, this has lead to highly efficient
event-scheduling strategies~\cite{Rapaport_1980,Isobe_1999}  and, for Monte
Carlo, to the development of cluster algorithms~\cite{Dress_2005, Jaster_1999,
Bernard_2009}. Even the modern simulation algorithm remain slow, however,
and revolutions like the cluster algorithms for spin
systems~\cite{Swendsen_1987,Wolff_1989} have failed to appear.
Moreover, rigorous mathematical bounds for the
correlation time (mixing time) of Monte Carlo algorithms were obtained in the
thermodynamic limit only for small densities~\cite{Kannan_2004,
Wilson_2000,Chanal_2010}, which are far inside the liquid phase. 
At higher densities, close to the liquid--solid transition, many numerical
calculations have suffered from insufficient simulation times until recently~\cite{Mak_2006,Bernard_2011}. 

In the present article, we discuss computational aspects of the
hard-disk model, starting with an introduction (\sect{algos}).
In particular, we reinterpret hard-sphere Monte Carlo
in terms of the sampling of points from  high-dimensional
polytopes (\sect{polytope}). Local Monte Carlo amounts to random walks in a
sequence of such polytopes, while event-chain Monte Carlo is equivalent to molecular dynamics
evolutions with particular initial conditions for the velocities. We analyze
the convergence properties of the algorithms in these polytopes for the
hard-disk case. Parallel event-chain algorithms emerge naturally as molecular
dynamics with more general initial conditions (\sect{parallel}). We describe
several parallelization strategies and report on implementations.

\section{Local Monte Carlo and event-chain Monte Carlo} 
\slabel{algos}

We consider $N$ equal hard disks of unit radius $\sigma=1$ in a square box of
size
$L\times L$. In the following, we assume
without mentioning periodic boundary conditions for positions and pair
distances. The statistical weights $\pi_a$ are equal to unity for configurations
$a$ without overlaps (all pair distances larger than $2$) and zero for
illegal configurations (with overlaps). The phase diagram of the system depends
only on the packing fraction $\eta := N\pi \sigma^2 / L^2$.  In the following, the
letters $a$, $b$, $c$, \dots,  label hard-disk configurations of $N$ disks,
given by the coordinates of the disk centers $\xvec_i$.  The letters $i$, $j$,
$k$ number disks.

\begin{figure}
\centering\includegraphics{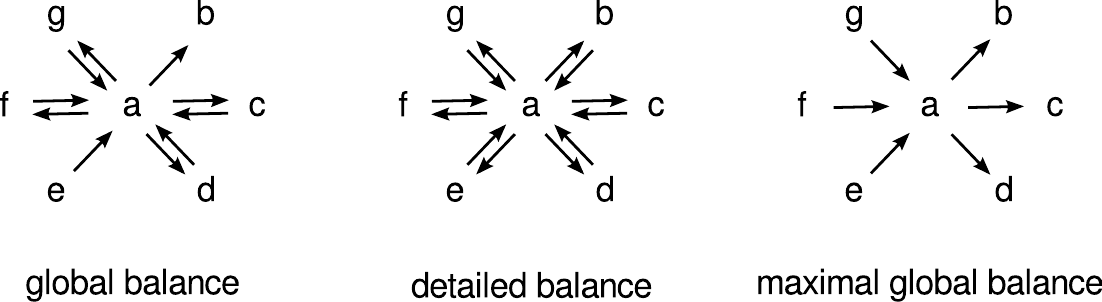}

\caption{Monte Carlo balance conditions:
Arrows represent probability flows $P_{a \to b} = \pi_a\; p_{a\to b}$ between configurations
$a$ and $b$ (each arrow stands for a probability flow of same magnitude).
\emph{Left}: Global balance, as required for Markov-chain Monte Carlo. The
total flow
$\sum_c P_{c \to a}$ into the configuration $a$ equals the flow $\sum_c P_{a
\to c}$ out of it. \emph{Center}: Detailed balance: the net flow between any two
configurations is zero, $P_{a \to b} = P_{b \to a}$.
\emph{Right}: Another special case of global balance: maximal global balance at 
$a$ ($P_{a \to b} > 0 \implies P_{b \to a} =  0 $).
\flabel{flows_detailed_global}
}
\end{figure}

\subsection{Balance conditions} 
\slabel{balances}
Markov-chain Monte Carlo algorithms are governed by balance
conditions for the flows $P_{a \to b} := \pi_a \; p_{a \to b}$ from
configuration $a$ to configuration $b$ (see \fig{flows_detailed_global}); $p_{a \to b}$ is
the conditional probability to move from $a$ to $b$, given that the 
system is in $a$. To converge towards the stationary distribution $\pi_a$, the
\emph{global} balance condition must be satisfied: The 
total flow onto configuration $a$ must equal the total flow out of $a$,
\begin{align}
   \sum_c P_{c \to a} =\sum_c P_{a \to c}.
\label{e:globalbal}
\end{align}

The local Monte Carlo algorithm, introduced by Metropolis et al. in 
1953~\cite{Metropolis_1953} (see \fig{event_chain_move}), uses the
more restrictive \emph{detailed} balance condition
$P_{a \to b} =  P_{b \to a}$ for which the net flow between each
pair of configurations $a$ and $b$  is zero.  Moving from configuration
$a = \{ \xvec_1, \ldots \xvec_i, \ldots, \xvec_N \}$
to
$b = \{ \xvec_1, \ldots \xvec_i + \deltavec, \ldots, \xvec_N \}$
involves sampling the disk $i$ to be displaced and the displacement
$\deltavec$.  For detailed balance, the probability to sample
$\deltavec$ at $\xvec_i$ must equal the probability to sample $-\deltavec$ at
position $\xvec_i'$.  
In order to be ergodic, the displacements $\deltavec$ are
chosen such that each disk can eventually reach any position in the system.

\cfigure[14cm]{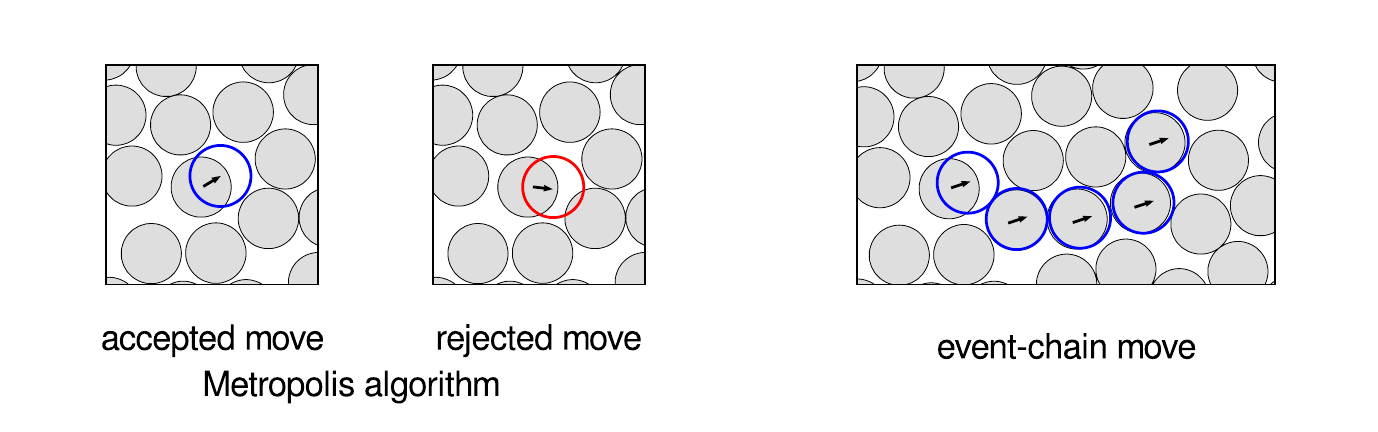}{Monte Carlo moves for hard disks.
\emph{Left}:  Accepted and rejected Metropolis
moves. \emph{Right}: Event-chain move. The sum of individual
displacements equals a predefined value $\ell$. With periodic boundary
conditions, the event-chain move is rejection-free. }

Unlike the local Monte Carlo algorithm, a single move of the event-chain
algorithm~\cite{Bernard_2009} may displace several disks.  An event-chain move
is parametrized by a total displacement $\ell$ and a direction $\elldir$,
which together form a vector $\ellvec := \ell\;\elldir$.  The move starts by
sampling a disk $i$ and \quot{sliding} it in the \elldir{} direction
until it hits another disk $j$, or at most for the
distance $\ell$.  The disk $j$ is then displaced in its turn, also in the
$\elldir$
direction, see \fig{event_chain_move}. This process continues until the
displacements of the individual disks sum up to $\ell$.  After this, a new disk
and possibly a new direction are sampled for the next move. With periodic
boundary conditions, no rejections occur in this algorithm.
For a given displacement vector $\ellvec$, any disk configuration $a$ can reach
$N$ other configurations, using each of the $N$ disks to start an event chain.
Likewise, $a$ can be reached from $N$ other configurations which
may be reconstructed by event chains with displacement vector
$-\ellvec$. This implies that the event-chain satisfies the global balance
condition, \eq{globalbal}.  If the vectors  $\pm \ellvec$ are equally likely,
it also satisfies detailed balance.
In order to be ergodic, the displacements $\ellvec$ must span space:
By choosing $\elldir\in\{\vec e_x, \vec e_y\}$,
the event-chain algorithm realizes the maximal global balance
(see \fig{flows_detailed_global}), where flow between two configurations is possible
only in one direction.  This version is more efficient than detailed balance versions
(for example, $\pm \vec e_x$ and $\pm \vec
e_y$)~\cite{Bernard_2009}.  It is again possible to alternate repeated moves
in the $\vec e_x$ direction with repeated moves in $\vec e_y$
without destroying the correctness of the algorithm.
For displacements $\ell$ smaller than the mean free path $l_{\rm mfp}$, the event-chain
algorithm is roughly equivalent to the local Monte Carlo algorithm. It
accelerates for increasing $\ell$, and for $\ell$ much larger than the mean
free path, it is about two orders of magnitude faster than the local Monte
Carlo method, and about ten times faster than the best current 
implementations~\cite{Isobe_1999} of event-driven molecular dynamics 
(see Ref.~\cite{Anderson_2012b}). 

\subsection{Correlation times and orientational order} 

\begin{figure}
\centering\includegraphics{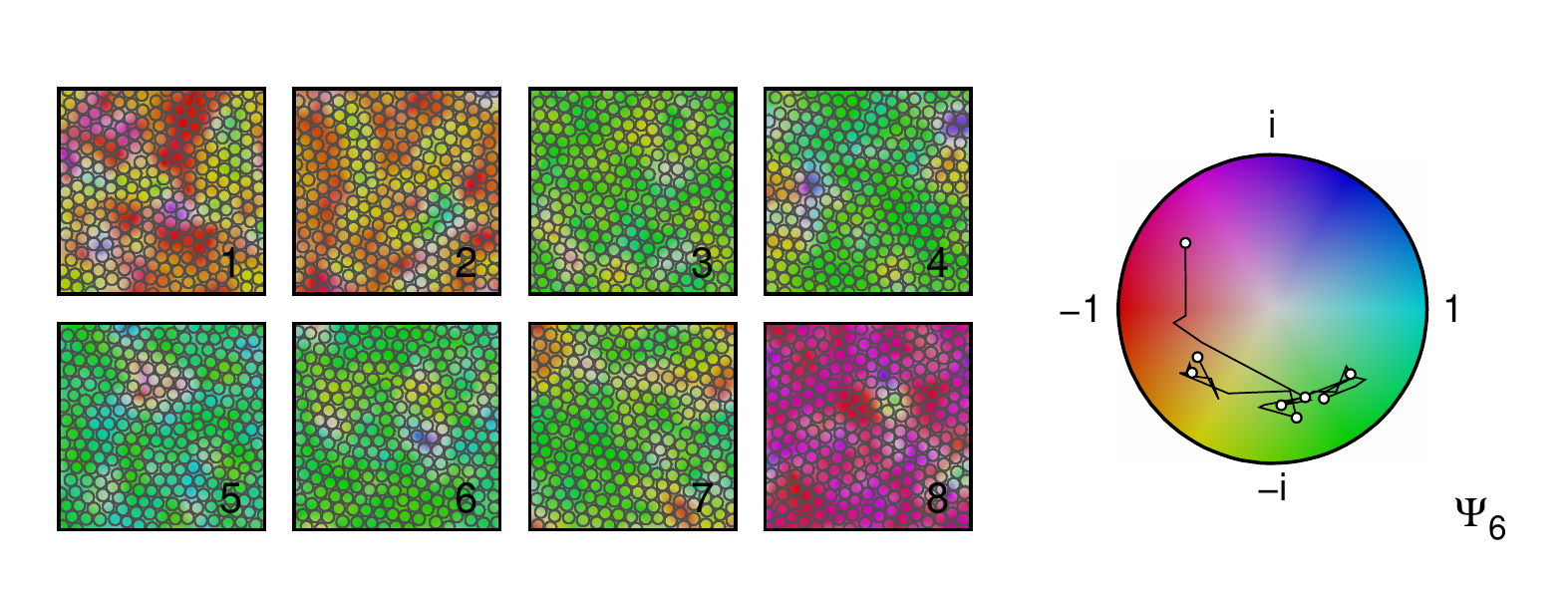}

\caption{Local Monte Carlo evolution
of $16^2$ disks in a square box with periodic boundary conditions
at packing fraction $\eta = 0.707$. \emph{Left}: Disk configurations
and their local orientational field $\psi_j$ for one simulation run.
Frame are separated by $100,000 N$ iterations ($10^5$ sweeps) of local Monte Carlo.
The slow decorrelation of the orientation is manifest.
\emph{Right}: Evolution of the global orientational order parameter $\Psi_6$,
\eq{global_psi}, in the complex plane, for the same simulation run.
\flabel{orientation_movie}}
\end{figure}

The characteristic challenge of numerical simulations for the hard-disk model
resides in the extremely long correlation time. This is illustrated in
\fig{orientation_movie} using snapshots of configurations obtained during a
long simulation run.  The system is quite small and not extremely dense, 
yet correlations in the orientation of the system persist over millions of Monte
Carlo moves. To quantify the orientations and their correlations, we consider
the local orientational field
\begin{equation}
  \psi_j := \sum_{k=1}^{N_j} w_{j,k} \exp(6{\rm i} \phi_{j,k}),
  \elabel{local_psi}
\end{equation}
where $N_j$ is the number of Voronoi neighbors of disk $j$. The $w_{j,k}$
(with $\sum_k w_{j,k} = 1$)
are normalized weights according to the length of the Voronoi interface
between disks $j$ and $k$, and $\phi_{j,k}$ is the angle of the vector
between the disk centers~\cite{MickelKapfer_2013}. The average of \eq{local_psi} over
all disks yields the global orientational order parameter,
\begin{equation}
  \Psi_6 := \frac{1}{N} \sum_j \psi_j.
  \elabel{global_psi}
\end{equation}
In a square box, the mean value of $\Psi_6$ is zero 
because of the $\phi_{j,k} \to \phi_{j,k}+\pi$ symmetry, and its
correlation function
\begin{equation}
  C_6(\Delta t) := \frac{\mean{\Psi_6(t) \Psi_6^*(t+
                \Delta t)}_t}{\mean{|\Psi_6(t)|^2}_t}.
\end{equation}
decays to zero for infinite times $\Delta t$. We conjecture that $\Psi_6$ is the
slowest observable in the system. For large times, global orientational
correlations decay exponentially, $C_6(\Delta t)\propto\exp(-\Delta t/\tau)$,
and we obtain the empirical correlation time $\tau$ from an exponential fit to
$C_6$.

\section{Polytope representation of event-chain moves}
\slabel{polytope}

\cfigure{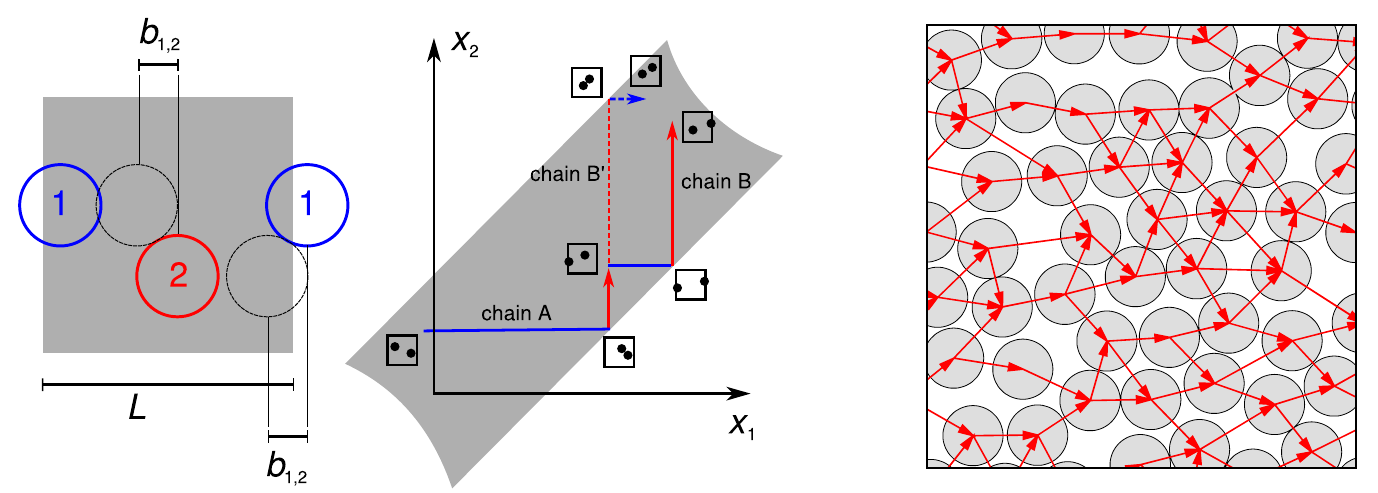}{Event-chain move and polytope
representation.  \emph{Left}: Two disks in a periodic box. The
constraints of \eq{polytopeEquations} are $x_{\rm 1} \leq x_{\rm 2} -
b_{\rm 1,2}$ and $x_{\rm 2} \leq x_{\rm 1} + L - b_{\rm 1,2} = x_{\rm 1} -
b_{\rm 2,1} $.  \emph{Center}: Molecular dynamics evolution in the polytope corresponding
to two event chains with moves of disk 1 (blue segments) and disk 2
(red segments). The trajectory begins with disk $1$, and it depends on the
choice of the starting disk ($1$ or $2$) for the second chain.
Periodic boundary conditions are ignored
for clarity. Snapshots of the configuration are sketched along the
trajectory.  \emph{Right}: Hard-disk configuration with its constraint graph
for motion along the $x$ axis. Each node has at most three forward and
three backward links. This graph is invariant under event-chain moves in
the $x$ direction.}

Event-chain moves along a single direction $\elldir=\ellvec / \ell$ 
sample a restricted configuration space.  For the remainder of this section,
we take the chains to move in the positive $x$ direction, unless specified
otherwise, to simplify the notation.
Since all $y$ coordinates are fixed, two disks whose $y$ coordinates differ by
less than $2$ radii
cannot slide across each other, and their relative order is fixed.
Furthermore, while in $x$ collision mode, any disk can collide with not more
than six other disks,
at most three in the forward direction, and at most three in backward direction (see 
\fig{polytope_ballistics_plus_graph}). The collision partners of a
disk may include itself, because of boundary conditions.  The relations among
disks constitute a \emph{constraint graph}, which expresses the partial
order between them (see \fig{polytope_ballistics_plus_graph}). This graph
remains invariant while performing event-chain moves in the $\pm \vec e_x$ direction.
Each directed edge from $i$ to $k$ corresponds to a linear inequality for the
$x$ coordinates of the disks $i$ and $k$:
\begin{align}
     x_i & \leq x_k - b_{i,k},
 \elabel{polytopeEquations} 
\end{align}
with $b_{i,k} := \sqrt{4 - (y_i-y_k)^2} $.
The constant $b_{i,k}$ can be adjusted to also account for periodic boundary
conditions in the $x$ direction.
The inequalities \eq{polytopeEquations} imply that no more than three
forward collision partners can be present\footnote{A superset of the actual
constraint graph can be computed from efficient local criteria.
This superset contains redundant inequalities, but describes the same polytope;
it is, in particular, useful for the practical implementation:
If $i$ collides forward with $j$, and $j$ collides forward
with $k$, we have $x_i \leq x_k - b_{i,j} - b_{j,k}$; if now $ b_{i,j} + b_{j,k}
> b_{i,k}$, the disks $i$ and $k$ can never come into contact; disk $j$ \emph{covers}
disk $k$.  Applying this rule iteratively, the disks in the forward direction
can be reduced to at most three: at most one each with $y\in (y_i-1,y_i+1)$,
with $y\in [y_i+1,y_i+2)$ and with $y\in (y_i-2,y_i-1]$.
Thus, each node in the constraint graph has at most degree six.
\label{computnote}
}.

The system of linear inequalities \eq{polytopeEquations} delimit a subset of the
$N$-dimensional
space of $x$ coordinates $\alltheX = (x_1, \dotsc, x_N$), 
an $N$-dimensional polytope, bounded by at most $3N$
hyperplanes. This convex
object is easier to analyze than the highly intricate $2N$-dimensional
configuration space of the full hard-disk problem. The polytope is
unbounded in the  $(1, 1, \dotsc,1)$ direction in consequence of the periodic
boundary
conditions, since uniform translation of all the disks is always permitted.
Also, since the constraint graph is invariant under event-chain moves
in the $x$ direction, so is the polytope.
However, the polytope becomes bounded by taking a section orthogonal to
$(1, 1, \dotsc,1)$.

In the invariant polytope, an event chain of total displacement
$\ell$ corresponds to a molecular dynamics evolution of duration
$\ell$: Displacing the $i$-th disk corresponds to the
``particle'' $\alltheX$ moving in the $i$-th coordinate direction,
and each collision event (the transfer of momentum from one disk to another) 
to a right-angle reflection at the facets of the polytope
(see \fig{polytope_ballistics_plus_graph}). The construction of the event-chain
move is finished at time $\ell$. The next move involves the sampling
of a new starting disk and possibly of one of the $\pm \elldir$ directions. 
In the invariant polytope, this is the choice of new velocities.
Local Monte Carlo on the other hand, if restricted
to moves in $x$ direction, implements diffusive motion in the invariant
polytope\footnote{To preserve the polytope, a \quot{sliding} version of local Monte Carlo
must be considered: The move $\xvec_i \to
\xvec_i + \deltavec$ is valid only if all intermediate positions
$\xvec_i + \alpha
\deltavec$ with $\alpha \in [0,1]$ yield legal hard-sphere configurations.}. 

The invariant constraint graph allows for fast lookup
of possible collision partners, and may even replace the customary cell
grids (see, for example, Section~2.4 of Ref.~\cite{SMAC}). While
computation of the actual constraint graph requires depth search, a superset
sufficient for practical computations can be computed efficiently, see the
footnote on page~\pageref{computnote}.

\subsection{Correlation functions in the invariant polytope} 

\begin{figure}
\includegraphics[trim=0 -2.7em 0 0,scale=.9]{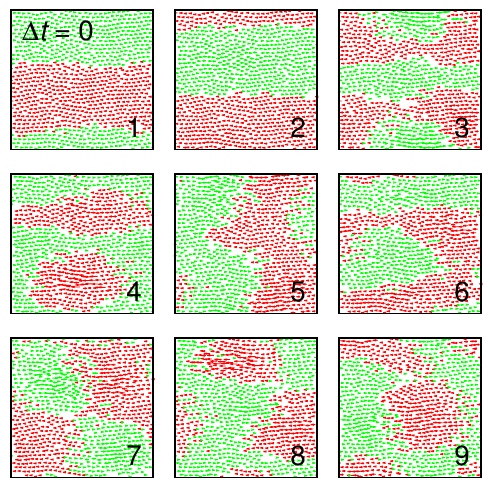}
\hspace{.2em}
\includegraphics[trim=0 -2.7em 0 0,scale=.9]{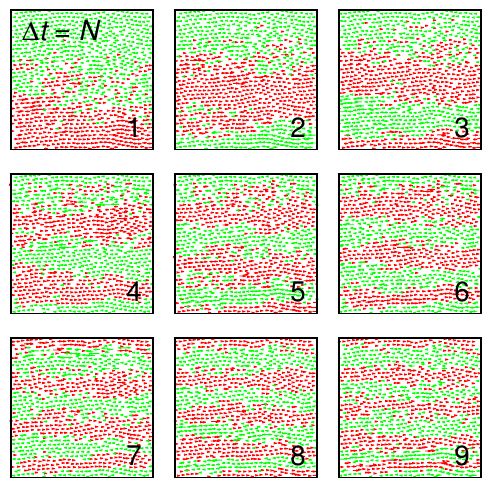}
\hspace{.2em}
\includegraphics{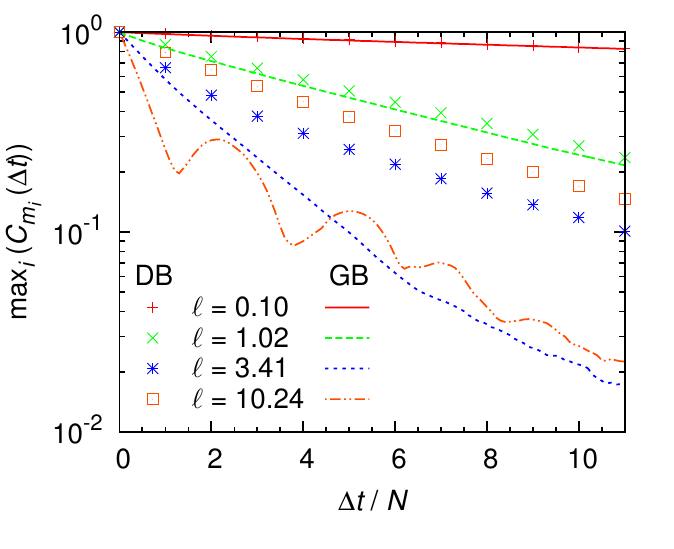}
\caption{
Relaxation dynamics in the invariant polytope, for a given initial configuration
of the $32^2$
hard-disk system at packing fraction $\eta= 0.698$. \emph{Left}:
Slowly decaying modes. Configurations are shown with 
red disks moving in the $+x$ direction and green disks in $-x$.
The modes shown are the eigenvectors of the largest eigenvalues of $U(0)$.
\emph{Center}: Remaining correlation after $\Delta t=N$ of event-chain moves in the
horizontal direction. These are the largest eigenvectors of $U(N)$. 
Correlations in the horizontal direction have all but disappeared.
\emph{Right}: Decay of the slowest modes, for the event-chain simulations
with various total displacements $\ell$, and for both the global (GB) and the detailed (DB)
balance version.\flabel{eigenmodes}
}
\end{figure}

Although the sampling problem from the invariant polytope concerns a convex
body, it is notoriously nontrivial~\cite{PolytopeComplexity}. 
The inequalities \eq{polytopeEquations}
essentially amount to a system of coupled one-dimensional hard-disk problems.
To study the relaxation behavior effected by the event-chain algorithm in the polytope,
we consider the cross-covariance of the disk coordinates,
\begin{align}
    U_{ij}(\Delta t) := \bigl\langle
            \tilde x_i(t+\Delta t) \cdot \tilde x_j(t)
        \bigr\rangle_t,
\end{align}
where $\tilde x_i(t)$ is the $x$ coordinate of the disk $i$, 
compensated for the overall translation of the system due to the
event-chain moves,
\begin{align}
    \tilde x_i(t) := x_i(t) - \frac{\alpha t}N - \left\langle x_i(t) -
            \frac{\alpha t}N \right\rangle_t.
\end{align}
Here, $\alpha$ is $1$ for the global balance version of the event-chain
algorithm (chains only in $+x$ direction), and $0$ for the detailed balance
version ($\pm x$).
The eigenvectors of $U(0)$ are the polytope's normal modes $m_i$, $i=1,\dotsc,N$,
in the sense of principal component analysis. The nature of the
modes $m_i$ depends on the structure of the invariant polytope and captures the
relative order of colliding disks and their frozen-in $y$ coordinates.  The
normal modes to the largest eigenvalues are large-scale cooperative rearrangements
of the disks (see \fig{eigenmodes}). They are the slowest modes to decay under
both local and event-chain Monte
Carlo and govern the global decorrelation of the disk configuration.  In particular,
two modes dominated by antiparallel flow bands are very slow to decay (mode 1 and
2 in \fig{eigenmodes}).

At delay times $\Delta t> 0$, the cross-covariance $U_{ij}(\Delta t)$
captures residual correlations among the disk coordinates. The event-chain
moves couple more efficiently to the longitudinal modes of the system, and we
find that after $\Delta t\approx N$, the event-chain algorithm has virtually 
erased longitudinal correlations. The most prominent residual
correlations carry a transverse band structure (see \fig{eigenmodes}).
The result is a substantial decrease in efficiency of the algorithm for
simulated duration in a single direction larger than $\approx N$.

To estimate the convergence time, we study the projection of the system's
evolution $\alltheX(t)$ onto a single mode, $\alltheX(t)\cdot m_i$.  The
autocorrelation function
\begin{align}
    C_{m_i}(\Delta t) := \frac{\langle \bigl( \alltheX(t)\cdot m_i\bigr)\bigl( \alltheX(t+\Delta t)\cdot m_i\bigr)\rangle_t}{\langle \bigl( \alltheX(t)\cdot m_i\bigr)^2 \rangle_t}
\end{align}
is, for short chain lengths $\ell$, monotonously decaying. Larger
chain lengths accelerate the decay, as the coupling
to large-scale modes is improved (\fig{eigenmodes}).  For chains spanning
several times the box, however, the autocorrelation functions
$C_{m_i}$ develop
oscillations with very weak damping, offsetting the benefits of
longer chains. The detailed balance version of event-chain
Monte Carlo is generally slower and less prone to oscillations. For optimal
performance, the global balance version should thus be used with $\ell$ larger,
but on the order of $l_{\rm mfp} \sqrt{N}$, and for times $\theta \approx N$
(see \fig{eigenmodes}).  For disk configurations larger than the correlation
length, $\ell$ can be reduced appropriately.

\subsection{Convergence of the full hard-disk problem}
\slabel{fullharddisk}

The invariant polytope representation allows us to interpret the convergence of
the full hard disk sampling
problem. The conceptually simplest Monte Carlo algorithm for hard disks
consists entirely in polytope sampling:  One iteration amounts to
direct sampling a new configuration $a_{n+1}$ from the invariant polytope of the
starting configuration $a_n$, and exchanging the $x$ and $y$ coordinates of
all the disks.  This Markov-chain algorithm satisfies detailed balance.
In our experiments, the timescale $\tau$, measured in iterations, for
relaxation to equilibrium increases only as
$N^{1/4}$ for large systems, implying that most of the complexity of the
hard-disk sampling problem resides in the polytope sampling.

Since direct sampling is a hard problem for high-dimensional polytopes
(see \sect{generalpolytope}), we replace it by Markov chains of a fixed number of
event-chain moves, in effect performing molecular dynamics in the invariant
polytopes for fixed duration $\theta$: 
\begin{equation}
\eqntext{polytope MD\\ in $x$ direction \\for duration $\theta$}
    \rightarrow
\eqntext{polytope MD\\ in $y$ direction \\for duration $\theta$}
    \rightarrow \cdots\;  . 
\end{equation}
This algorithm satisfies detailed or global balance depending on the version of
the event-chain algorithm that is used for polytope sampling.

\begin{figure}
\centering
\includegraphics{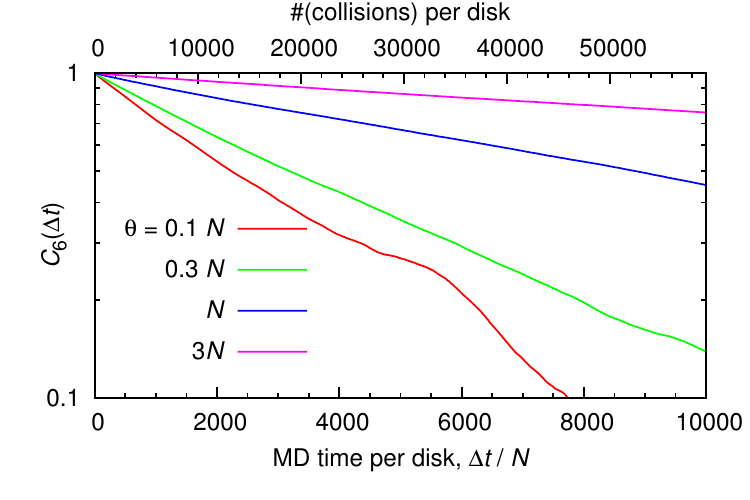}
\includegraphics{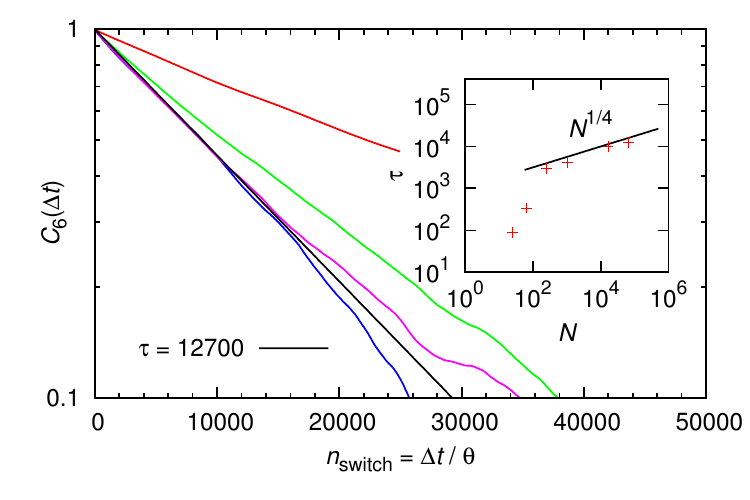}
\caption{\emph{Left}: Decay of the $\Psi_6$ autocorrelation function
$C_6(\Delta t)$ for several switching intervals $\theta$,
as a function of the simulated MD time (\emph{bottom axis}), or
alternatively, the number of collisions per disk (\emph{top axis}).
\emph{Right}:  Decay of $C_6$ as a function of the number of $x$/$y$ switches $n_{\text{switch}}$.
As $\theta$ approaches $N$, the curves approach the limit of direct sampling
from the polytope, with a mixing time of $\tau$ cycles.
All curves were averaged from systems of $N=256^2$ disks at packing fraction
$\eta=0.698$; the chain length was $\ell=6.5\cdot 10^3$.  \emph{Inset}:
The mixing time $\tau$ first increases
rapidly with system size, but only grows as $N^{1/4}$
for larger systems (also $\eta=0.698$). }
\label{f:switchingAndMixing}
\end{figure}

We study the influence of the switching interval $\theta$ on convergence
properties.  In \fig{switchingAndMixing}, the autocorrelation function $C_6$ of
the complex order parameter $\Psi_6$ is plotted vs.~cumulative molecular
dynamics time. $C_6$
decays most quickly when the switching interval $\theta$ is small, but the decay
speed deteriorates very slowly with $\theta$.  Only at $\theta\approx N$
(corresponding to about $6$-$7$ collisions per disk at these densities), the
algorithm becomes notably less efficient.  The efficiency drop thus follows the
decay of longitudinal (in $x$ direction) correlations in the
invariant polytope, and is to be expected
from the results in \sect{polytope}.

In the limit $\theta\rightarrow\infty$, the event-chain algorithm realizes 
direct sampling in the invariant polytope. The approach to this limit
is illustrated in \fig{switchingAndMixing} by plotting $C_6$ against
the number of x/y switching cycles $n_{\text{switch}}$.  As the switching interval $\theta$
increases, the autocorrelation functions approach an asymptotic curve
$\propto\exp(-n_{\text{switch}}/\tau)$, where $\tau$ is the correlation time
of the direct sampling algorithm.  We find that for practical purposes,
event-chain Monte Carlo reaches the asymptotic regime for $\theta\approx N$,
and thus samples an approximately independent point in the invariant polytope
in $O(N)$ operations.  Importantly, the correlation time $\tau(N)$ increases
rapidly only for small system size $N$. After the system size surpasses the 
correlation length, $\tau$ grows only as $N^{1/4}$.

\subsection{Application to general polytopes}
\slabel{generalpolytope} 

The invariant polytope is bounded by hyperplanes which are normal to $N-2$
coordinate axes and have unit derivative along the remaining axes.  By choice
of the $\ellvec$, the molecular dynamics evolution is aligned with the coordinate axes at all
times, and computations of intersections are of complexity $O(1)$.  As shown
in \sect{fullharddisk}, the event-chain algorithm seems to achieve
an effective mixing time of $O(N)$ collision events, so that the
cost of sampling the hard-disk polytope appears as $O(N)$.

The event-chain algorithm also
allows to sample general polytopes. Direct sampling from polytopes is
straightforward only in low dimensions $N$, especially in $N=2$: A two-dimensional
polytope with $n$ edges (a convex $n$-sided polygon), can be decomposed into
$n$ triangles, using an interior point. Triangles may then be sampled according
to their areas, and a random point may be sampled inside the sampled triangle
(see, e.\,g.~chap. 6.2 of~\cite{SMAC}). In higher dimensions $N$, triangulation
by simplices generalizes this decomposition. Since polytopes such as
the invariant hard-disk polytope have an exponential number of facets, direct
sampling algorithms are no longer practical. Markov-chain 
sampling~\cite{Smith_1984,Rubin_1984,Dyer_1991,Kannan_2012} achieves mixing
times of $O(M N)$ steps, where $M$ is the number of bounding hyperplanes ($M\leq
3N$ for hard disks), and where each move may be implemented in $O(M)$ steps. 
It will be interesting to see how event-chain polytope sampling compares with
existing polytope sampling methods, in particular the `hit-and-run' algorithms.

\section{Parallel Monte Carlo algorithms for hard disks} 
\slabel{parallel}

In view of the long running times of Monte Carlo simulations and of the
current standstill in computer clock speeds, it is essential
to develop \emph{parallel Monte
Carlo methods} which distribute the work load among several
\emph{threads} performing independent computation with as few communication as
possible. Such
methods will allow to study not only the standard hard disk ensemble, but also
related systems such as soft disks and polydisperse disk packings.  However,
parallel Monte Carlo algorithms for continuum systems pose many more problems
than for lattice models, for example the Ising spins, where straightforward
parallel application of local Metropolis updates converges to the Boltzmann
distribution~\cite{Berg_2004}.

\subsection{Parallel implementation of local Monte Carlo } 

A massively parallel implementation of the local Monte Carlo
algorithm was applied recently to the hard-disk melting 
problem~\cite{Anderson_2012a,Anderson_2012b}. It sets up square cells according
to a
four-color checkerboard pattern. Disks in same-color cells
can be updated simultaneously, but moves across cell boundaries are rejected.
To ensure ergodicity, a new cell grid must be sampled periodically. 
Massive parallelism of $\sim 1500$ threads on a graphics card offsets the
slowness of local Monte Carlo compared to event-chain algorithm Monte Carlo~\cite{Anderson_2012b}.
These calculations confirmed the first-order liquid-hexatic phase transition
in hard disks~\cite{Bernard_2011}.

\begin{figure}
\centering
\includegraphics[height=2.2in,trim=0 -.2cm 0 0]{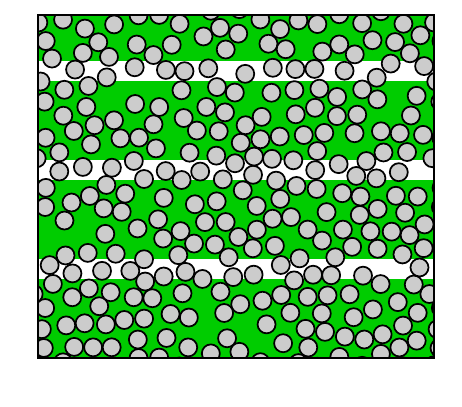}
\includegraphics{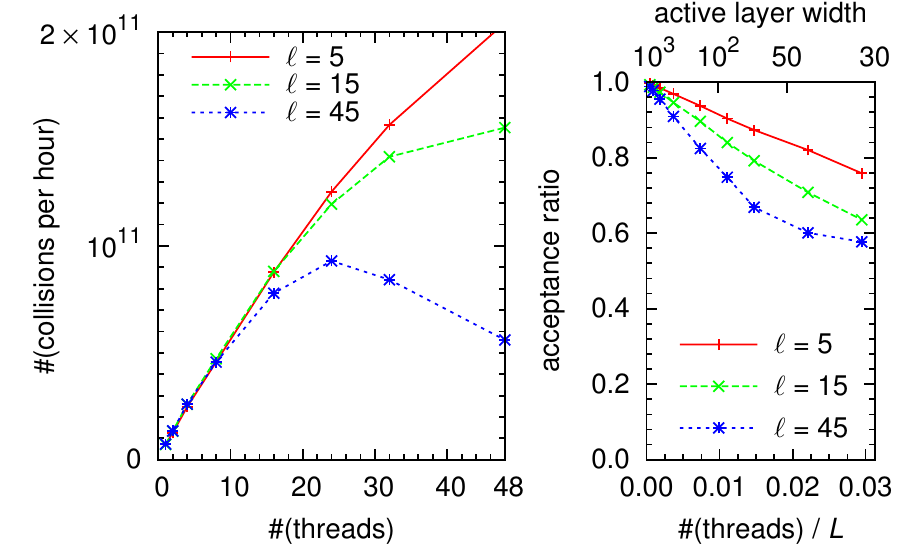}

\caption{
\emph{Left}: Two-color stripe scheme with active (green backdrop) and
isolation layers for the event-chain algorithm. Chains may run simultaneously if
they are located in different stripes.
\emph{Center}: Scaling of the isolation layer algorithm on a shared-memory
machine (Opteron 6276, 2.3 GHz), for a $N=4096^2$ disk packing at $\eta=0.698$. We plot the number of
collisions in accepted chains per hour of computation. On the same machine,
the serial version has a performance of about $8.5\cdot 10^{9}$ collisions per hour.
The event-chain routine is the same in both programs.
\emph{Right}:  The acceptance ratio for chains depends on the thickness of the
active layers (which is decreases as more threads are added) and the total displacement $\ell$ of the chains.
\flabel{stripescheme}
}
\end{figure}

\subsection{Parallel implementations of event-chain Monte Carlo}

For parallel implementations of event-chain Monte Carlo, we consider only
parallel threads that run chains in the same direction $\pm \elldir$. This 
minimizes the chance that two chains cross each other 
and move the same disks. It also allows us to apply the invariant polytope
framework of \sect{polytope}.
It is instructive to realize that the effects of an event-chain move
can be summarized in the difference vector of the new and old $x$ coordinates:
$\Delta\alltheX = \alltheX(\ell) - \alltheX(0)$, with $\alltheX(t) = (x_1(t), \dotsc,
x_N(t))$.
Moreover, if two chains are \emph{independent}, meaning their sets of disks touched
are disjoint, the net effect of running both chains is the sum of their individual difference vectors,
$\Delta\alltheX^{\rm net}=\Delta\alltheX^{(1)}+\Delta\alltheX^{(2)}$. If, however,
any disk is touched by both chains, the chain reaching this disk earlier in
MD time has precedence, the later chain sees a modified environment, and consequently takes
a different evolution.  Thus, interdependent chains cannot be added
arithmetically\footnote{Note, however, that due to the convexity of the
accessible configuration space, the arithmetic average of two moves,
$(\Delta\alltheX^{(1)}+\Delta\alltheX^{(2)} )/2 $, is always admissible.}.
The primary obstacle in parallelizing event-chain Monte Carlo consists in
preserving the correct causal relations between subsequent chains, as required
for the convergence to correct equilibrium distribution.

In the following, we discuss three strategies to parallelize event-chain Monte
Carlo.  The \emph{predict/execute algorithm} distributes work among threads
for a model of chains that follow each other
chronologically. The effects $\Delta\alltheX$ of several chains are predicted in
advance from the current disk configuration.
The effects of the chains are then applied to the system state in
the chronological order in which the starting disks were sampled.
To detect conflicts, it is sufficient to compute the intersection of the set of
disks touched by the current chain and of the chains that ran since the
beginning of planning; if this intersection is not empty, the
chain has to be recomputed from the updated state of the disk configuration.
Planning and execution of chains can proceed in parallel on a
shared-memory machine. Using lock-free
data structures, we attain collision rates in excess of $10^{11}$ per hour in
$x$ collision mode on a four-processor machine.  Due to its serial nature, this
algorithm does not scale well beyond a few threads, however; with too many
chains predicted in advance, the probability for recomputations rises. Moreover,
switching between $x$ and $y$ collision modes requires reinitialization of the
data structures and is rather expensive.

A variation of the four-color scheme adapted to the event-chain algorithm
partitions the system in horizontal stripes, separated by
\emph{frozen isolation layers} of thickness $\geq 2$ disk radii (see \fig{stripescheme}).
Disks with their centers in the isolation layers are kept fixed, and thus guarantee
the independence of chains running in neighboring stripes.  To
preserve the isolation layers, chains colliding with a frozen disk are rejected.
As there are rejected moves, the global balance condition is no longer guaranteed:
the number of accepted forward chains can be different
from the number of accepted backward chains (see \sect{balances}).  When allowing chains in both
the $\pm\elldir$ directions, however, the isolation layer algorithm satisfies
detailed balance.  Furthermore, in order to limit the rejection rate,
the per-chain total displacement $\ell$ has to be kept lower than in the serial
algorithm.  In view of the discussion of \sect{polytope}, these
necessities reduce somewhat the efficiency of the method.  Due to the isolation
layers, the accessible configuration space is restricted, and
for ergodicity, the layer boundaries have to be resampled
periodically, as in the four-color version of local Monte Carlo.

We have implemented the isolation layer algorithm in parallel on
a shared-memory machine.  Using several cores in parallel, it is possible
to achieve effective collision rates which are 10--30 times the
single-core performance (see \fig{stripescheme}), for systems of sufficient
size.  For systems too small, less threads can be used without shrinking
the active strips to a point where the acceptance ratio becomes a limiting factor.
Systems of physical interest, however, are on the order of $N=1024^2$, 
and allow to use 10--20 cores with moderate $\ell$.  At this time, the
algorithm is not bound by rejection rates, but by communication
between threads.

Finally, the event-chain scheme is not fundamentally limited to a single
moving disk at any time.  We may indeed launch multiple \emph{concurrent chains},
which run at the same simulated MD time, and interact with each other.  This is
different from the parallel simulation of chains which interact in sequential
manner. In the invariant polytope picture, multiple concurrent chains correspond
to choosing more general initial conditions, where more than one disk
is given an initial velocity of $1$.  After time $\ell$, multiple chains
have executed, and possibly interacted with each other; there is no rejection
in this algorithm.  The problem has some resemblance with
event-driven molecular dynamics, because the scheduling of collisions must be foreseen,
but there are several simplifications: all velocities are in the same direction
and of magnitude $0$ or $1$. As a consequence, two moving disks
cannot collide with one other; however, the faithful simulation
of chains close by and possibly interacting requires careful
synchronization among threads. In our experiments, this  limits the speedup by
parallelization. Our most efficient method at this point is the isolation layer
algorithm.

\section{Conclusion}

We have reached in this paper a better understanding of the event-chain Monte Carlo
algorithm for the hard-disk sampling problem.  By restricting the algorithm to
chains in a single direction, a connection appears to the well-known problem of
sampling random points from a polytope:  A move of the event-chain algorithm
consists in performing a finite-time molecular dynamics simulation in the invariant
polytope of the disk configuration. This connection offers new strategies to
solve the hard-disk sampling problem in terms of polytope sampling; it also
suggests to investigate the utility of event-chain methods for the sampling of
general polytopes.  Finally, it will be interesting to study the combinatorial
structure of the typical invariant polytope, and its dependence on
thermodynamical parameters.

By the study of correlation functions, we have shown that the Monte Carlo 
relaxation process in the invariant polytope separates into two phases:
A rapid longitudinal
relaxation, followed by a much slower relaxation of the transverse degrees of
freedom.  We have given recommendations for the parameters of the algorithm
based on these results.  Finally, we have discussed several strategies for
parallelizing
Monte Carlo algorithms for hard disks, alleviating the problem of the long simulation
times in hard disk Monte Carlo.  The parallelization of the hard-disk ensemble
remains challenging due to its unique combination of very little actual computation
and long correlation times.  Efficient methods to tackle the hard-disk ensemble are,
however, crucial in order to treat related systems such as soft disks with the same
level of success as the hard disks.  New concepts such as the link to polytope
sampling will be essential in this effort.

\section*{Acknowledgment}
We thank P. Diaconis, E. P. Bernard, S. Leitmann and M. Hoffmann for fruitful discussions.

\end{document}